\documentclass[aps,pra,letterpaper,superscriptaddress,twocolumn,floatfix]{revtex4-2}
\usepackage {graphicx}
\usepackage{amsmath}
\usepackage[usenames, dvipsnames]{color}

\allowdisplaybreaks

\begin{document}

\title{Phase Transition of the Ising Model on a 3-Dimensional Fractal Lattice}

\author{Jozef \textsc{Genzor}}
\email[]{jozef.genzor@gmail.com}
\affiliation{Institute of Physics, Slovak Academy of Sciences, D\'ubravsk\'a cesta 9, 84511 Bratislava, Slovakia}
\affiliation{Department of Physics, Graduate School of Science, Kobe University, Kobe 657-8501, Japan}

\author{Roman \textsc{Krčmár}}
\affiliation{Institute of Physics, Slovak Academy of Sciences, D\'ubravsk\'a cesta 9, 84511 Bratislava, Slovakia}

\author{Hiroshi \textsc{Ueda}}
\affiliation{%
  Center for Quantum Information and Quantum Biology,
  The University of Osaka, 1-2 Machikaneyama, Toyonaka 560-0043, Japan
}%
\affiliation{
  RIKEN Center for Computational Science (R-CCS),
  Kobe, Hyogo 650-0047, Japan
}

\author{Denis \textsc{Kochan}}
\affiliation{Institute of Physics, Slovak Academy of Sciences, D\'ubravsk\'a cesta 9, 84511 Bratislava, Slovakia}
\affiliation{Department of Physics and Center for Quantum Frontiers of Research and Technology (QFort), National Cheng Kung University, Tainan 70101, Taiwan}

\author{Andrej \textsc{Gendiar}}
\affiliation{Institute of Physics, Slovak Academy of Sciences, D\'ubravsk\'a cesta 9, 84511 Bratislava, Slovakia}

\author{Tomotoshi \textsc{Nishino}}
\affiliation{Department of Physics, Graduate School of Science, Kobe University, Kobe 657-8501, Japan}

\date{\today}

\begin{abstract}
The critical behavior of the classical Ising model on a three-dimensional fractal lattice with Hausdorff dimension $d_H = \ln32 / \ln4 = 2.5$ is investigated using the higher-order tensor renormalization group (HOTRG) method. We determine the critical temperature $T_c \approx 2.65231$ and the critical exponents for magnetization $\beta \approx 0.059$ and field response $\delta \approx 35$. Unlike a previously studied 2D fractal with $d_H \approx 1.792$, the specific heat for this 3D fractal exhibits a divergent singularity at $T_c$. The results are compared with those for regular lattices and other fractal structures to elucidate the role of dimensionality in critical phenomena.
\end{abstract}

\maketitle

\section{Introduction}

Dimensionality is a cornerstone in the theory of phase transitions and critical phenomena \cite{Domb1972Phase, Fisher1974Critical}. For systems on regular lattices, which possess translational invariance, the spatial dimension $d$ critically influences the universality class of the transition. However, many systems in nature and in artificial constructs exhibit fractal geometries \cite{Mandelbrot1982}. These structures are characterized by self-similarity over a range of scales, a lack of translational invariance, and often non-integer Hausdorff dimensions ($d_H$). Such characteristics pose profound questions regarding the conventional understanding of critical behavior and the very notion of dimensionality that governs it. For fractal systems, besides $d_H$, other dimensional measures such as the spectral dimension $d_s$ \cite{Alexander1982}, the order of ramification $R$ \cite{Gefen1983CriticalG}, or dimensions derived from boundary scaling (e.g., $M \sim L^{d-1}$ for $M$ boundary sites of a cluster of size $L$) can be defined. A central, unresolved issue is which of these, if any, serves as the effective dimension dictating critical exponents and the validity of hyperscaling relations \citep{Stauffer1994, Stanley1971}.

Sierpiński gaskets are canonical examples of finitely ramified fractals. The classical Ising model on the 2D Sierpiński gasket ($d_H = \ln3/\ln2 \approx 1.585$), as well as its higher-dimensional generalizations, famously shows no phase transition at finite temperature \cite{Gefen1980Critical, Luscombe1985Critical}, despite $d_H > 1$. This is often linked to its finite order of ramification, implying a quasi-one-dimensional character at large scales. In contrast, infinitely ramified fractals like Sierpiński carpets can sustain phase transitions \cite{Monceau1998}. The advent of Tensor Network (TN) algorithms, particularly the Higher-Order Tensor Renormalization Group (HOTRG) \cite{Xie2012}, has provided robust numerical tools to tackle phase transitions on self-similar fractal lattices by efficiently managing their recursive structure. Our group has leveraged HOTRG to explore classical Ising models on various 2D fractals, including a 4x4 base fractal ($d_H \approx 1.792$) \cite{2dising, ad}, a 6x6 base fractal ($d_H \approx 1.934$) \cite{acta}, a $J_1$-$J_2$ tunable fractal connecting this 4x4 fractal to a regular square lattice \cite{j1j2}, and the Sierpiński carpet ($d_H \approx 1.893$) \cite{sierp}. Quantum phase transitions on 2D Sierpiński fractals have also been studied by us \cite{qsierp}.

\begin{figure}[htbp]
\includegraphics[width=0.45 \textwidth]{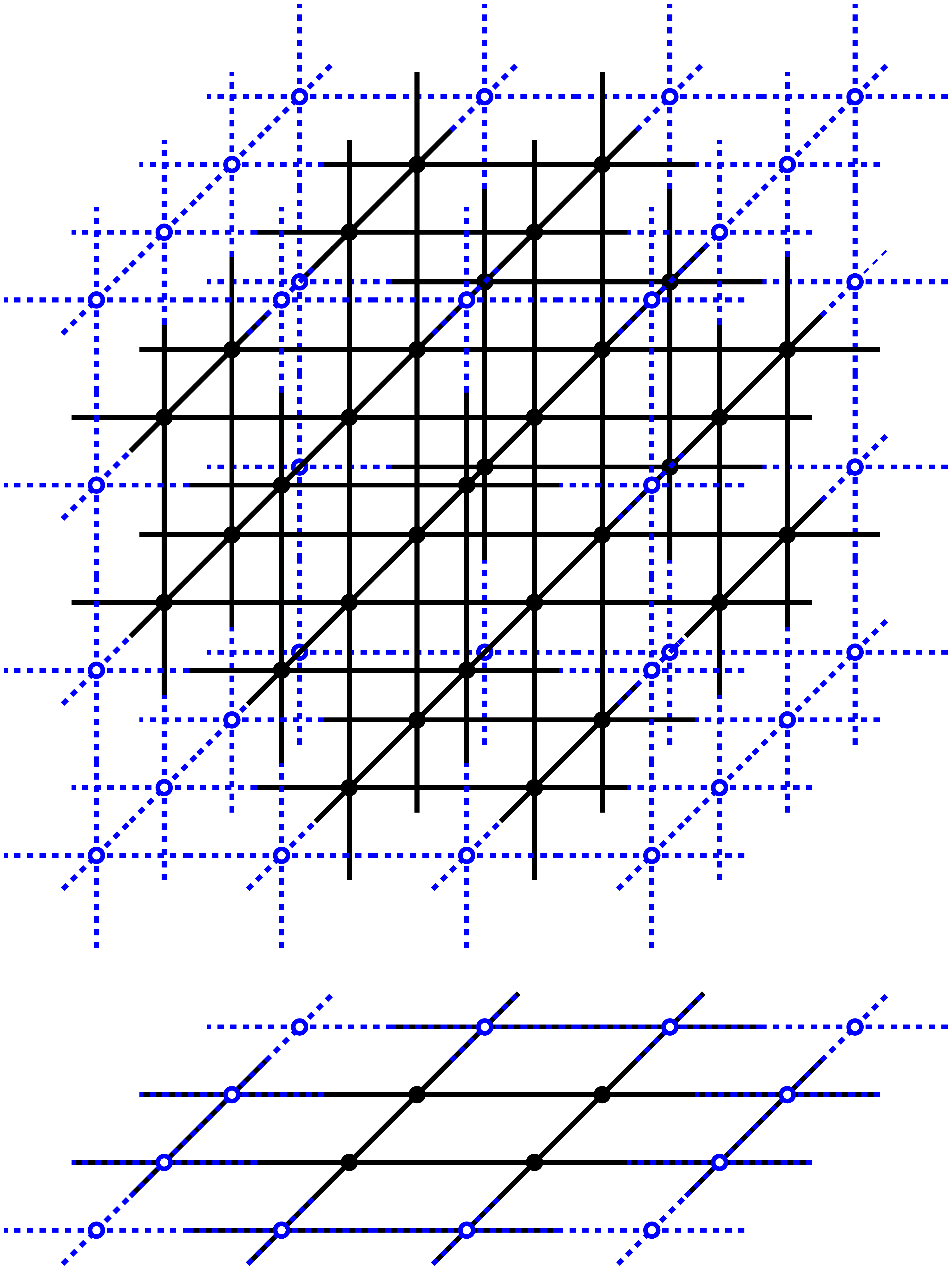}
\caption{
(Color online)
Upper part: Composition of the fractal lattice.
The basic cluster which contains $N_{n=1} = 32$ vertices in the first iteration step.
In each step $n$ of the system extension, the linear size of the system increases by a factor of 4, where only 32 units (full circles) are linked (solid lines), and where the remaining 32 units (empty circles, dashed lines) at the edges of the $4\times4\times4$ cube are missing.
Lower part: A projection from one of the six sides of the basic cluster.
}
\label{fig:Fig_1}
\end{figure}

This paper extends our previous investigations to a three-dimensional (3D) fractal lattice. The lattice is constructed by a recursive extension process where, at each step $n$, the linear size of the system increases by a factor of four. A basic unit is composed of 32 smaller, identical units, while 32 units are absent from the corners and edges of a full $4 \times 4 \times 4$ regular cubic arrangement (see Fig.~\ref{fig:Fig_1}). Consequently, the number of sites scales as $N_n = 32^n$, yielding a Hausdorff dimension $d_H = \ln32 / \ln4 = 2.5$. In contrast, the number of outgoing bonds from a cluster increases by a factor of four in each step. This implies a dimension derived from boundary scaling, $M \sim L^{d-1}$, of $d=2$. This juxtaposition of $d_H=2.5$ with an effective boundary dimension $d=2$ motivates a central question. Our prior work on a 2D fractal with $d_H \approx 1.792$ (and $d=1.5$) found no divergence in the specific heat \cite{2dising}. Given that the regular 2D square lattice ($d_H=2, d=2$) exhibits a logarithmic specific heat singularity \cite{Onsager1944}, it is pertinent to ask whether the current 3D fractal ($d_H=2.5, d=2$) will show a specific heat divergence.

The primary objective of this study is to investigate the phase transition of the classical Ising model on this novel 3D fractal lattice using the HOTRG method. We aim to determine its critical temperature $T_c$ and the critical exponents $\beta$ (associated with spontaneous magnetization) and $\delta$ (characterizing the critical isotherm). A particular focus will be on the nature of the specific heat. By comparing our findings with established results for regular lattices and other fractals, we seek to further elucidate the intricate relationship between varied dimensional characteristics and critical phenomena on non-homogeneous systems.

The paper is structured as follows: Section~\ref{sec:ModelRepresentation} details the model and its tensor-network representation. Our numerical results are presented and analyzed in Section~\ref{sec:NumericalResults}. Finally, Section~\ref{sec:Conclusions} offers a summary, discussion, and concluding remarks.

\section{Model representation}  
\label{sec:ModelRepresentation}

Tensor-network representation can be employed for any classical statistical system with local (i.e. short-range) interactions. 
The Hamiltonian of the Ising model is defined as
\begin{equation}
{\cal H} = - J \sum_{\left<i j\right>}^{~} \sigma_i \sigma_j - h \sum_{i} \sigma_i  \, ,
\end{equation}
where $\sigma$ takes either $+1$ or $-1$, $J > 0$ represents the ferromagnetic coupling, and $h$ is a constant external magnetic field imposed to each spin. For simplicity of the further explanation, let us consider the case with no external field by setting $h=0$.  
The partition function of the Ising model defined on the 3D fractal lattice can be expressed in terms of tensor network states defined by five types of local tensors represented by $T$, $X$, $Y$, $Z$ and $Q$,
\begin{eqnarray} 
\label{eq:T_def}
T_{x_i^{~} x'_i y_i^{~} y'_i z_i^{~} z'_i} &=& \raisebox{-1.6em}{\includegraphics[height=3.5em]{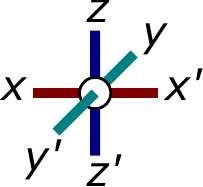}} \\
 &=& \displaystyle\sum_{\xi}^{~} W_{\xi x_i^{~}} W_{\xi x'_i} W_{\xi y_i^{~}} W_{\xi y'_i} W_{\xi z_i^{~}} W_{\xi z'_i} \, , \nonumber \\
\label{eq:X_def}
X_{x_i^{~} x'_i y_i^{~} z_i^{~}} &=& \raisebox{-1.6em}{\includegraphics[height=3.5em]{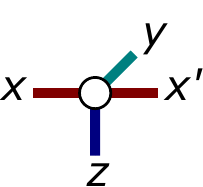}} \\
&=& \displaystyle\sum_{\xi}^{~} W_{\xi x_i^{~}} W_{\xi x'_i} W_{\xi y_i^{~}} W_{\xi z_i^{~}} \, , \nonumber \\
\label{eq:Y_def}
Y_{y_i^{~} y'_i x_i^{~} z_i^{~}} &=& \raisebox{-1.6em}{\includegraphics[height=3.5em]{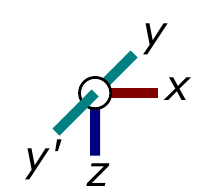}} \\
&=& \displaystyle\sum_{\xi}^{~} W_{\xi y_i^{~}} W_{\xi y'_i} W_{\xi x_i^{~}} W_{\xi z_i^{~}} \, , \nonumber \\
\label{eq:Z_def}
Z_{z_i^{~} z'_i y_i^{~} x_i^{~}} &=& \raisebox{-1.6em}{\includegraphics[height=3.5em]{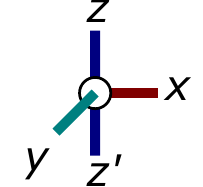}} \\
&=& \displaystyle\sum_{\xi}^{~} W_{\xi z_i^{~}} W_{\xi z'_i} W_{\xi y_i^{~}} W_{\xi x_i^{~}} \, , \nonumber \\
\label{eq:Q_def}
Q_{x_i^{~} y_i^{~} z_i^{~}} &=& \raisebox{-1.6em}{\includegraphics[height=3.5em]{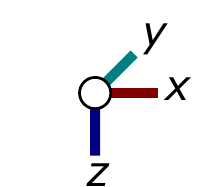}} = \displaystyle\sum_{\xi}^{~} W_{\xi x_i^{~}} W_{\xi y_i^{~}} W_{\xi z_i^{~}} \, , 
\end{eqnarray}
where $W$ is a $2\times2$ matrix determined by the bond weight factorization. 
While the choice for $W$ is arbitrary to a certain degree, here we choose an asymmetric factorization
\begin{equation} \label{weight}
W = \left(\begin{array}{lr} 
\sqrt{\cosh 1 / T} & \sqrt{\sinh 1 / T}   \\
\sqrt{\cosh 1 / T} & -\sqrt{\sinh 1 / T} \end{array} \right) \, ,
\end{equation}
where $T$ is the temperature. 
Notice that when two local tensors are contracted via non-physical (auxiliary) index $x$, the bond weight ${\cal{W}}_{\rm B} \left(\sigma_{i}, \sigma_{j}\right) = \exp{\left(\sigma_{i} \sigma_{j} / T \right)}$ is correctly re-expressed
\begin{equation}
{\cal{W}}_{\rm B} \left(\sigma_{i}, \sigma_{j}\right) = \sum_{x=0}^{1} W_{\xi_i x} W_{\xi_j x} \, ,
\end{equation}
where the first matrix index $\xi_i = (1 - \sigma_i) / 2$ takes values of $0$ and $1$ when $\sigma_i = 1$ and $\sigma_i = -1$, respectively.
For clarity in the following tensor-network diagrams, we use different colors to distinguish the $x$, $y$, and $z$ directions.
A coarse-graining renormalization procedure is used to calculate the partition function. 
We start counting the iteration steps from zero; therefore, we denote the initial tensors in Eqs.~\eqref{eq:T_def}--\eqref{eq:Q_def} as $T^{(n=0)}=T$, $X^{(n=0)}=X$, $Y^{(n=0)}=Y$, $Z^{(n=0)}=Z$, and $Q^{(n=0)}=Q$. 
At each iterative step $n$, the new tensors $T^{(n+1)}$, $X^{(n+1)}$, $Y^{(n+1)}$, $Z^{(n+1)}$, and $Q^{(n+1)}$ are created from the previous-iteration tensors $T^{(n)}$, $X^{(n)}$, $Y^{(n)}$, $Z^{(n)}$, and $Q^{(n)}$.
Practically, this is achieved in several steps.
First, we construct the ``core'' tensor $S^{(n)}$ by contracting eight copies of tensor $T^{(n)}$ into a $2\times2\times2$ cube.
The ``core'' tensor $S^{(n)}$ can be found in the center of new tensors $T^{(n+1)}$, $X^{(n+1)}$, $Y^{(n+1)}$, $Z^{(n+1)}$, and $Q^{(n+1)}$.
Depending on what type of tensor is constructed, different legs with two composed indices ($L^{(n)}_{[X]}$, $L^{(n)}_{[Y]}$, and $L^{(n)}_{[Z]}$) or spikes with one composed index ($C^{(n)}_{[X]}$, $C^{(n)}_{[Y]}$, and $C^{(n)}_{[Z]}$) are attached to the corresponding sides of the ``core'' tensor.
By repeating this procedure, we can construct a lattice structure as large as required.
We will now explain this construction in detail.

The ``core'' tensor $S^{(n)}$ is constructed by contracting eight copies of tensor $T^{(n)}$
\begin{align} 
\label{S_def}
&S^{(n)}_{ \substack{
(x_1^{~} x_2^{~} x_3^{~} x_4^{~}) (x'_1 x'_2 x'_3 x'_4) (y_1^{~} y_2^{~} y_3^{~} y_4^{~}) \\
(y'_1 y'_2 y'_3 y'_4) (z_1^{~} z_2^{~} z_3^{~} z_4^{~}) (z'_1 z'_2 z'_3 z'_4)
} } \nonumber \\
 &\quad=\quad \raisebox{-2.1em}{\includegraphics[height=4.2em]{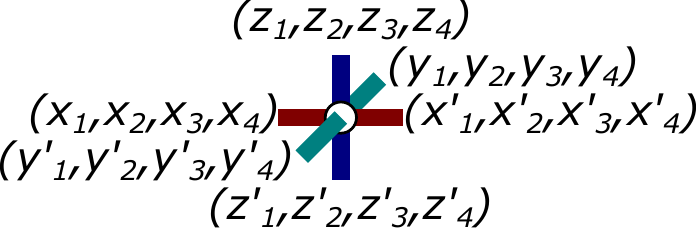}} \nonumber \\
 &\quad=\quad \raisebox{-4.25em}{\includegraphics[height=8.5em]{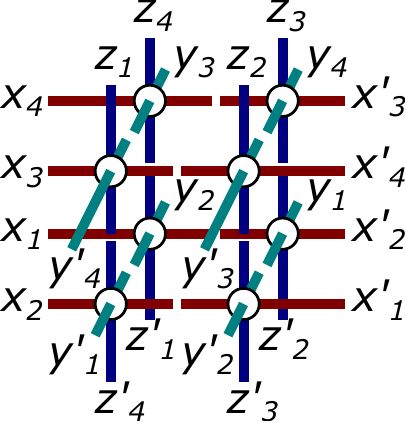}} \, .
\end{align}
This is achieved by performing three HOTRG extension steps (i.~e., in all three directions) as in the case of a regular 3D lattice.
We depict the tensor $S^{(n)}$ using thick lines. 
The leg tensor $L^{(n)}_{[X]}$ is constructed from four copies of tensor $X^{(n)}$, $L^{(n)}_{[Y]}$ from four copies of $Y^{(n)}$, and $L^{(n)}_{[Z]}$ from four copies of $Z^{(n)}$ as follows
\begin{align} 
\label{eq:LX_def}
&L^{(n)}_{[X] (x_1^{~} x_2^{~} x_3^{~} x_4^{~}) (x'_1 x'_2 x'_3 x'_4)} \nonumber \\
 &\quad=\quad \raisebox{-2.1em}{\includegraphics[height=4.2em]{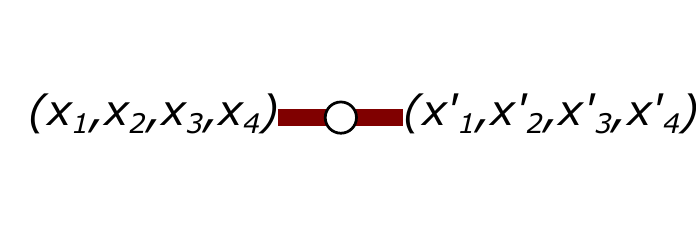}} \nonumber \\[-2.0em]
 &\quad=\quad \raisebox{-4.25em}{\includegraphics[height=8.5em]{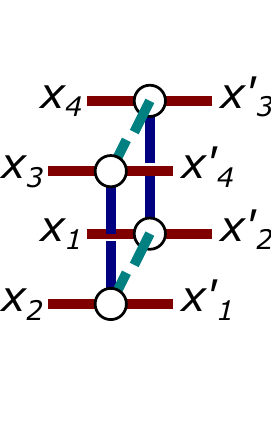}} \, , \\
\label{eq:LY_def}
&L^{(n)}_{[Y] (y_1^{~} y_2^{~} y_3^{~} y_4^{~}) (y'_1 y'_2 y'_3 y'_4)} \nonumber \\
 &\quad=\quad \raisebox{-2.1em}{\includegraphics[height=4.2em]{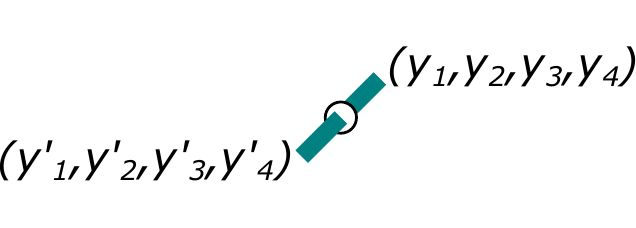}} \nonumber \\[-1.0em]
 &\quad=\quad \raisebox{-4.25em}{\includegraphics[height=8.5em]{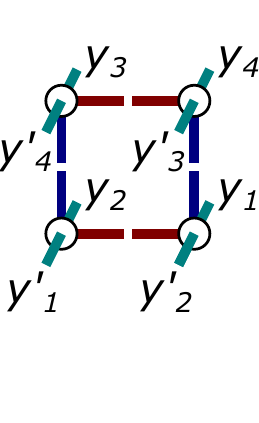}} \, , \\
\label{eq:LZ_def}
&L^{(n)}_{[Z] (z_1^{~} z_2^{~} z_3^{~} z_4^{~}) (z'_1 z'_2 z'_3 z'_4)} \nonumber \\
 &\quad=\quad \raisebox{-2.1em}{\includegraphics[height=4.2em]{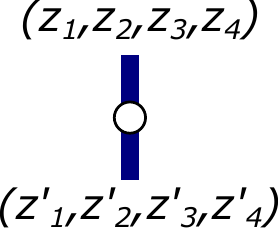}} \nonumber \\[-2.0em]
 &\quad=\quad \raisebox{-2.75em}{\includegraphics[height=8.5em]{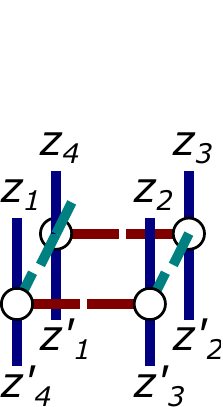}} \, .
\end{align}
The spike tensors $C^{(n)}_{[X]}$, $C^{(n)}_{[Y]}$, and $C^{(n)}_{[Z]}$ are all constructed from four copies of tensor $Q^{(n)}$ rotated and permuted correspondingly as follows
\begin{align} 
\label{eq:CX_def}
&C^{(n)}_{[X] (x_1^{~} x_2^{~} x_3^{~} x_4^{~})} \nonumber \\
 &\quad=\quad \raisebox{-2.1em}{\includegraphics[height=4.2em]{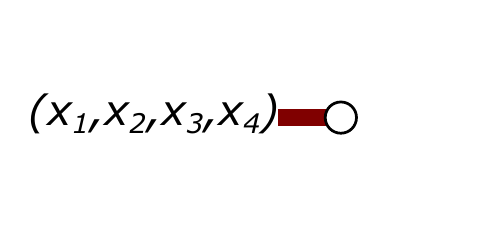}} \nonumber \\[-2.0em]
 &\quad=\quad \raisebox{-4.25em}{\includegraphics[height=8.5em]{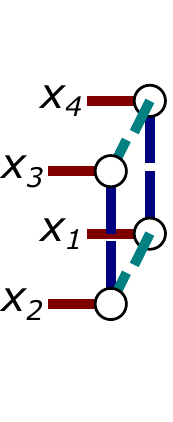}} \, , \\
\label{eq:CY_def}
&C^{(n)}_{[Y] (y_1^{~} y_2^{~} y_3^{~} y_4^{~})} \nonumber \\
 &\quad=\quad \raisebox{-2.1em}{\includegraphics[height=4.2em]{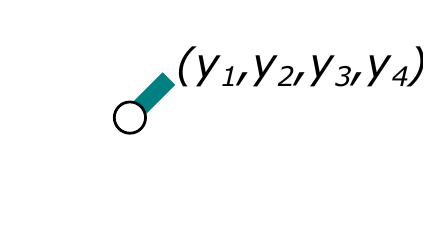}} \nonumber \\[-1.0em]
 &\quad=\quad \raisebox{-5.25em}{\includegraphics[height=8.5em]{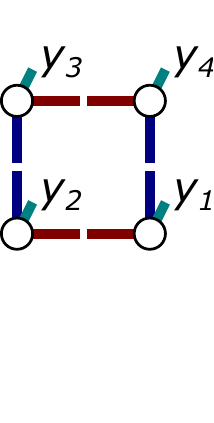}} \, , \\[-1.0em]
\label{eq:CZ_def}
&C^{(n)}_{[Z] (z_1^{~} z_2^{~} z_3^{~} z_4^{~})} \nonumber \\
 &\quad=\quad \raisebox{-2.5em}{\includegraphics[height=4.2em]{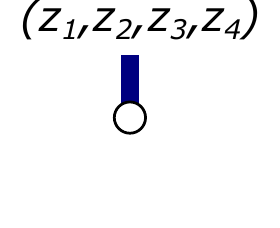}} \nonumber \\[-3.0em]
 &\quad=\quad \raisebox{-2.75em}{\includegraphics[height=8.5em]{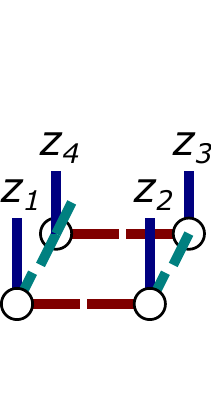}} \, .
\end{align}
With all auxiliary objects prepared (central tensor, legs, and spikes), we are ready to create new tensors $T^{(n+1)}$, $X^{(n+1)}$, $Y^{(n+1)}$, $Z^{(n+1)}$, and $Q^{(n+1)}$ for the next iteration step $n+1$.
The local tensors are updated as follows
\begin{align} 
\label{eq:T_new_def}
T_{x x' y y' z z'}^{(n + 1)} &=& \raisebox{-3.8em}{\includegraphics[height=8em]{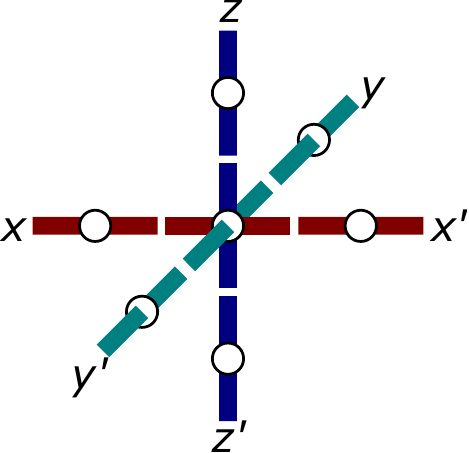}} \, , \\
\label{eq:X_new_def}
X_{x x' y z}^{(n + 1)} &=& \raisebox{-3.8em}{\includegraphics[height=8em]{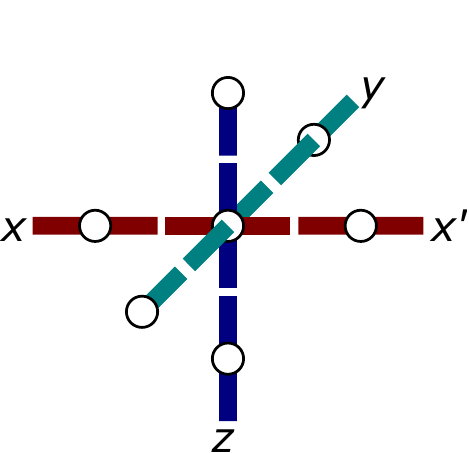}} \, , \\
\label{eq:Y_new_def}
Y_{y y' x z}^{(n + 1)} &=& \raisebox{-3.8em}{\includegraphics[height=8em]{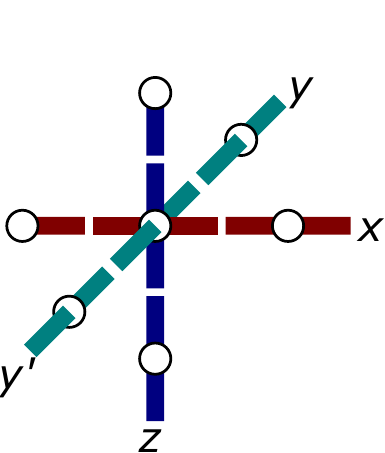}} \, , \\[1.0em]
\label{eq:Z_new_def}
Z_{z z' y x}^{(n + 1)} &=& \raisebox{-3.8em}{\includegraphics[height=8em]{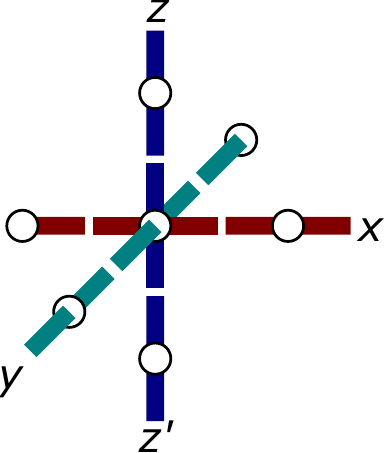}} \, , \\
\label{eq:Q_new_def}
Q_{x y z}^{(n + 1)} &=& \raisebox{-3.8em}{\includegraphics[height=8em]{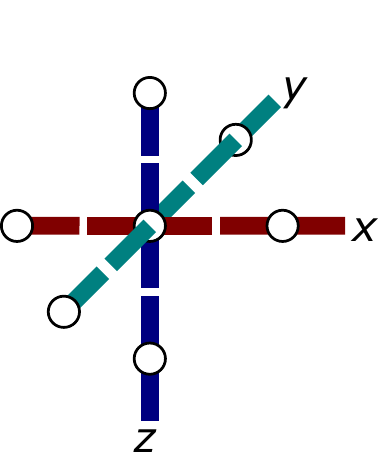}} \, . 
\end{align}
The partition function $Z_n(T)$ of the system after $n$ extensions is evaluated as
\begin{equation}
Z_n(T) = \sum_{ijk} T^{n}_{i i j j k k } \, ,
\end{equation}
where we impose the periodic boundary conditions.

\section{Numerical Results}
\label{sec:NumericalResults}

We set the ferromagnetic coupling $J=1$ and work in units where the Boltzmann constant $k_{\mathrm{B}}=1$. 
In the numerical calculation by means of HOTRG, we keep $D = 18$ states at most for block spin variables.

\begin{figure}[htbp]
\includegraphics[width=0.45 \textwidth]{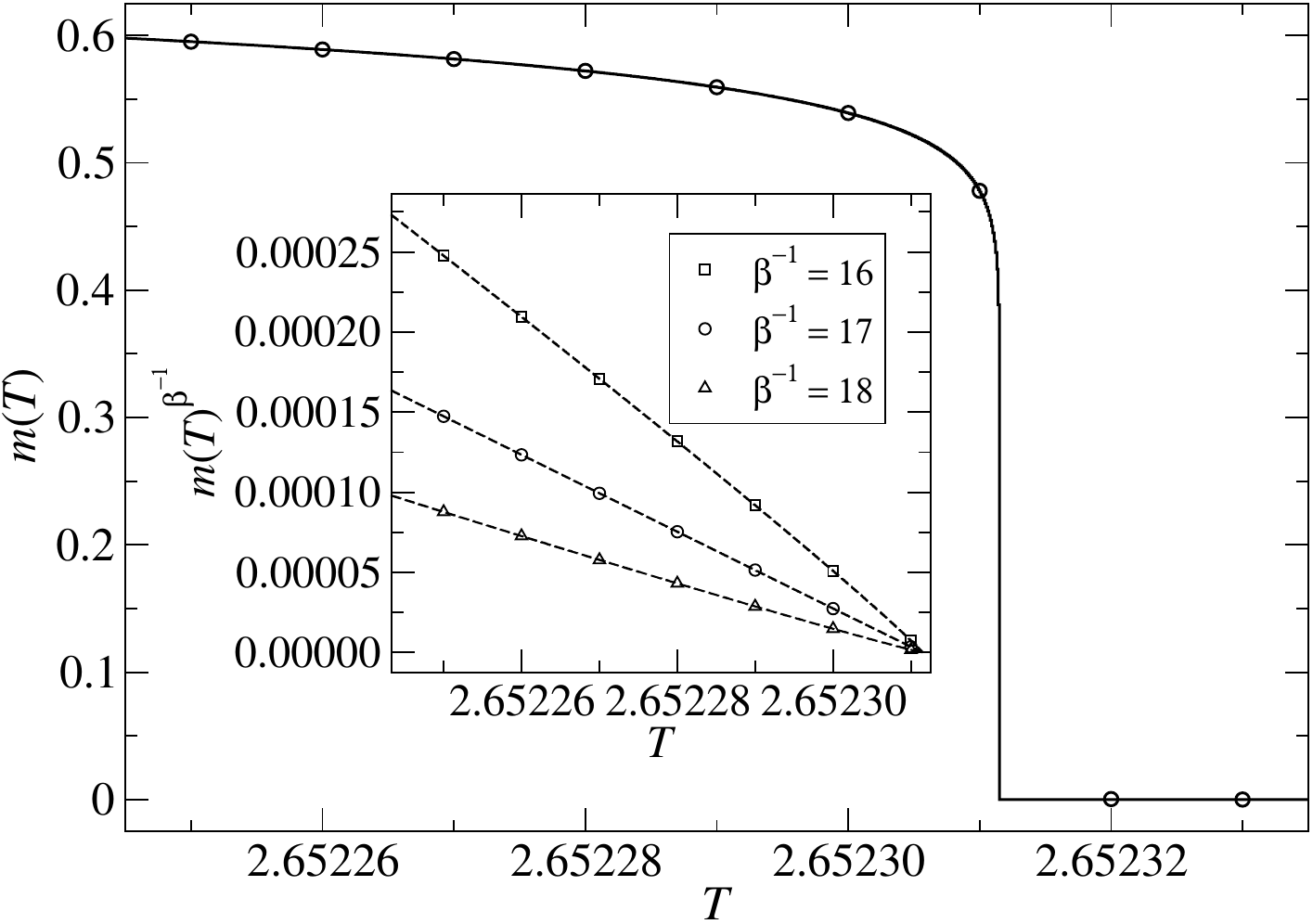}
\caption{
The spontaneous magnetization per site $m(T)$ ($D=18$). Inset: the power-law behavior below $T_{\text{c}}=2.65231$.
}
\label{fig:Fig_2}
\end{figure}

Let us first look at the spontaneous magnetization obtained by introducing 
an impurity to the system. A single impurity is located somewhere in the ``central'' area of the lattice 
(the system is rotated after every HOTRG extension step
in order to keep the impurity close to the ``center'' of the lattice).
Figure~2 shows the spontaneous megnetization $m(T)$ of the system. 
Right below the critical temperature $T_{\text{c}}=2.65231$, 
the magnetization exhibits the typical power-law behavior
\begin{equation}
m( T ) \sim 
| T_{\rm C}^{~} - T |^{0.059}_{~} \, . 
\end{equation}
The critical temperature $T_{\text{c}}$ is determined with high precision by tracking its convergence with respect to the bond dimension $D$, as shown in Fig.~3. For $D \ge 16$, the value is stable to six significant digits.
The precision of the exponent might be inferred from the inset of the Fig.~2 
(as a small deviation from the linear dependence) as well as from the inset of the Fig.~3 
(as a small fluctuation with respect to the bond dimension $D$). 
The relative difference between $\beta (D=17)$ and $\beta (D=18)$ is $0.2\%$. 
\textbf{Notice that even though the currently studied fractal is three-dimensional, 
the critical exponent $\beta \approx 0.059 \approx 1/17$ is smaller than the exponent in the case of the square-lattice 
Ising model $\beta_{\text{square}}=1/8$}. 

\begin{figure}[htbp]
\includegraphics[width=0.45 \textwidth]{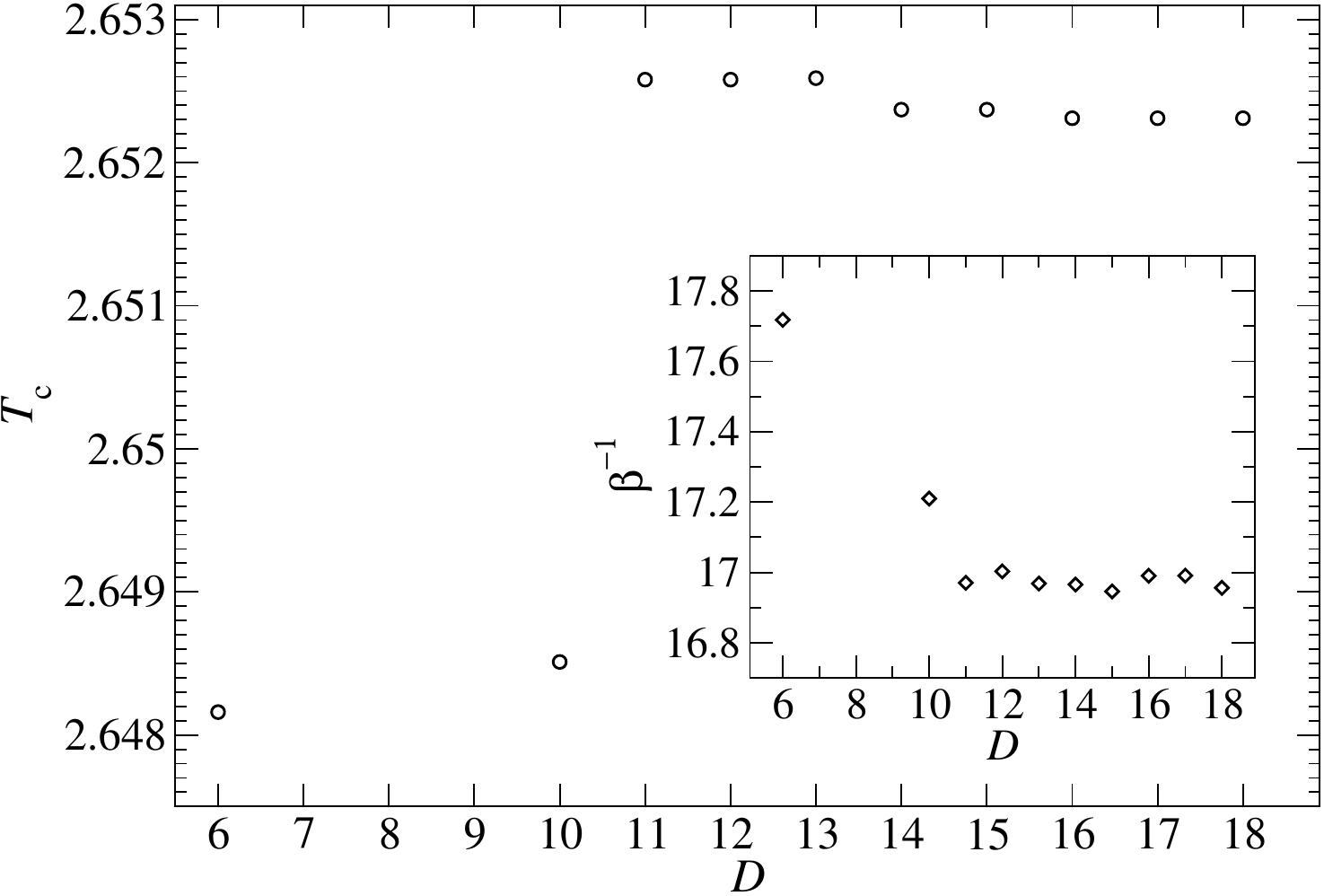}
\caption{
The convergence of the critical temperature $T_{\text{c}}$ with respect to the bond dimension $D$. 
For $D=\{16,17,18\}$, the critical temperature $T_{\text{c}}=2.65231$ is identical up to 6 digits. 
Inset: The convergence of the $\beta$ exponent (here, the inverse values $\beta(D)^{-1}$ are presented) with respect to $D$. 
The cases of $D=\{7,8,9\}$ are degenerated and therefore not included in the plots. 
}
\label{fig:Fig_3}
\end{figure}

The numerical value of the critical exponent $\beta$ is weakly dependent on the exact location of the imputity tensor. 
By averaging the impurities positioned on eight central tensors of the basic cluster, we have obtained slightly different 
values than showed above; $\beta \approx 0.066 \approx 1/15$ for D=17, 
whereas the relative difference between $\beta (D=16)$ and $\beta (D=17)$ is less than $10^{-4}$. 

\begin{figure}[htbp]
\includegraphics[width=0.45 \textwidth]{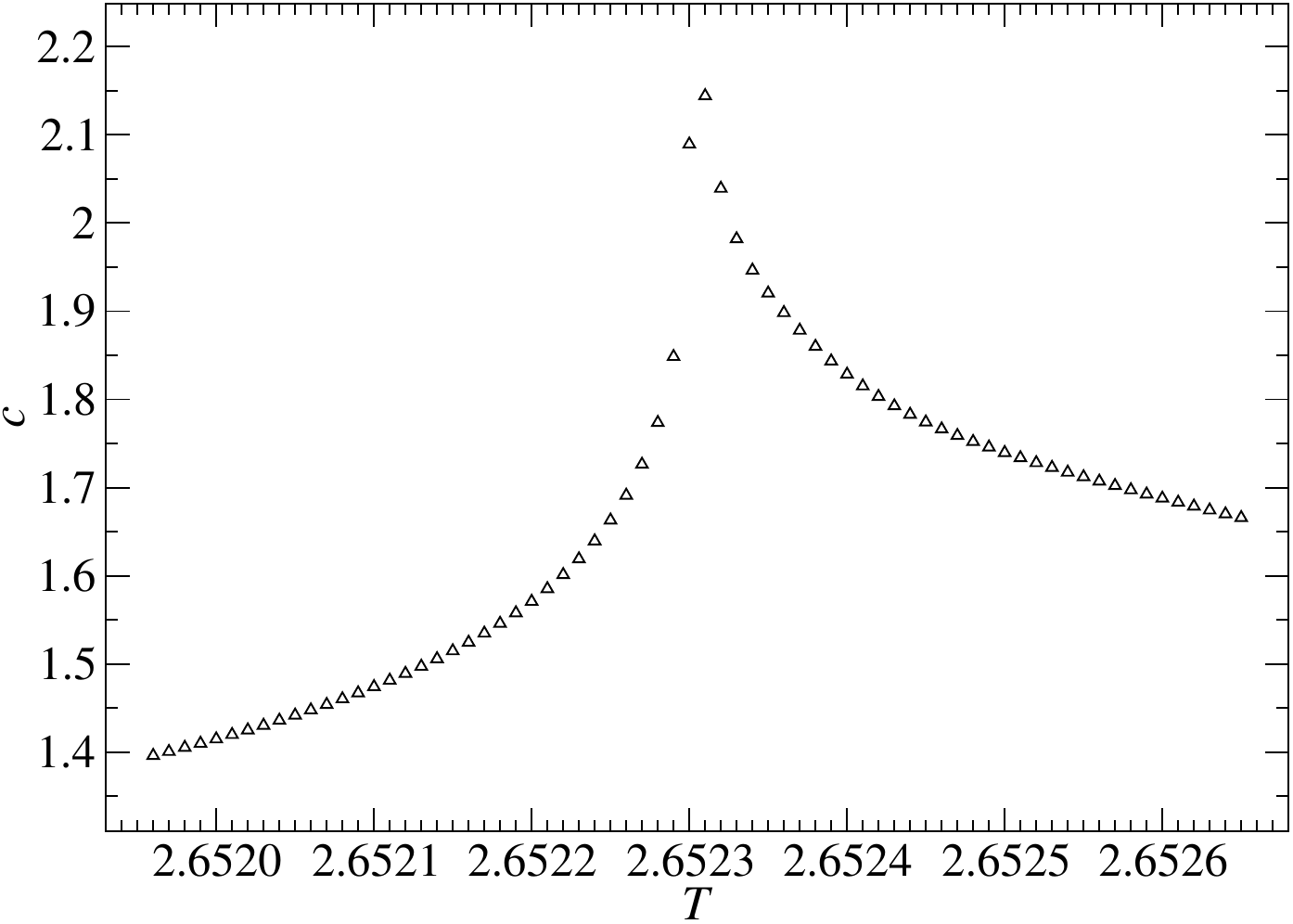}
\caption{
The specific heat $c(T)$ per site (obtained from the data for the bond energy for $D=16$ as a numerical derivative with respect to $T$). 
Notice that the specific heat $c(T)$ diverges around $T\approx2.6523$, 
which is in an agreement with the $T_{\text{c}}=2.65231$ determined from the magnetization.  
}
\label{fig:Fig_4}
\end{figure}

We have observed a singular behavior (divergence) at $T_{\text{c}}$ of the specific heat per site $c(T)$ 
obtained as a numerical derivative of the bond energy $u$ 
with respect to the temperature, i.~e., $c(T)=d u/ dT$, see Fig.~4. 
We would like to emphasise that \textbf{this behavior is different from one observed in the case of 
the 2-dimensional fractal lattice studied earlier~\cite{2dising}}, where no divergence was found. 

\begin{figure}[htbp]
\includegraphics[width=0.45 \textwidth]{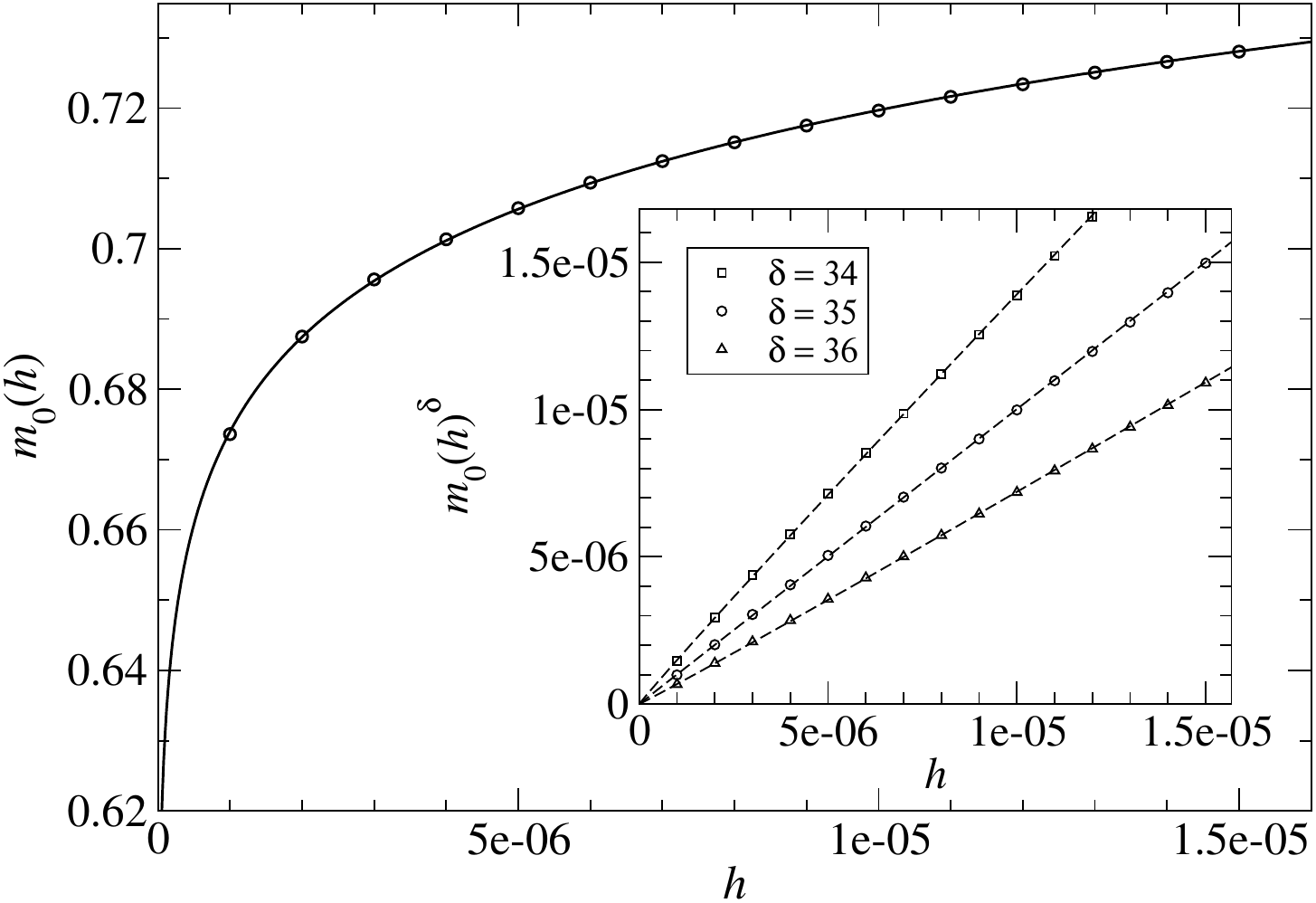}
\caption{
The magnetic field response of the spontaneous magnetization $m_{\text{0}}=m(T=T_{\text{c}})$ at
the temperature $T_{\text{c}}=2.65231$ ($D=18$). 
Inset: the power-law dependence of $m_{\text{0}}$ on the non-zero external field $h$. 
}
\label{fig:Fig_5}
\end{figure}

In order to determine another independent critical exponent, we have also numerically examined the magnetic 
field response of the spontaneous magnetization $m$ at $T_{\text{c}}$, see Fig.~5. 
\begin{equation}
m_{0}^{}(h)=m( T=T_{\text{c}}, h ) \sim h^{1/35}_{~} \, . 
\end{equation}
The relative difference between $\delta (D=17)$ and $\delta (D=18)$ is $0.4\%$. 

Let us now consider the averaged magnetization to obtain the critical exponent $\delta$.  
As before, we have obtained slightly different 
values than showed above; $\delta \approx 31$ for D=17, 
whereas the relative difference between $\delta (D=16)$ and $\delta (D=17)$ is again very small, i.~e., less than $10^{-4}$. 

We also attempted to determine the critical exponent $\alpha$, which governs the singular behavior of the specific heat near the critical temperature via the relation $c(T) \sim |T - T_c|^{-\alpha}$. While our analysis of the specific heat data (Fig.~4) consistently points to a small, positive value for $\alpha$ (in the range of $0.05$ to $0.07$), the weak nature of this singularity makes a more precise determination challenging with the available numerical accuracy.

\section{Conclusions and Discussions}
\label{sec:Conclusions}

In this work, we have investigated the critical behavior of the classical Ising model on a novel three-dimensional fractal lattice using the higher-order tensor renormalization group (HOTRG) algorithm. The lattice is characterized by a Hausdorff dimension $d_H = 2.5$ and a boundary scaling dimension $d=2$. Our numerical simulations precisely determined the critical temperature to be $T_c \approx 2.65231$. We calculated the critical exponent for spontaneous magnetization as $\beta \approx 0.059 \approx 1/17$ and the critical isotherm exponent as $\delta \approx 35$. These findings provide new insights into the complex relationship between geometry and critical phenomena in systems lacking translational invariance.

A principal finding of this study is the observation of a divergent singularity in the specific heat at $T_c$. This result, when placed in the context of previous findings, helps to clarify the conditions under which such a singularity emerges on fractal lattices. We can observe a clear progression: while the 4x4 fractal ($d_H \approx 1.792$, $d=1.5$) shows no divergence \cite{2dising}, both the 6x6 fractal ($d_H \approx 1.934$, $d \approx 1.774$) \cite{acta} and our current 3D fractal ($d_H=2.5$, $d=2$) do. This ordered sequence strongly suggests that the onset of a divergent specific heat is governed by a dimensional threshold. Our results, combined with prior work, help to bracket this threshold between the dimensional characteristics of the 4x4 and 6x6 fractals. The presence of a divergence in our case, where $d=2$, draws a compelling parallel to the regular 2D square lattice, which famously exhibits a logarithmic singularity \cite{Onsager1944} and reinforces the idea that system connectivity is a key factor.

The calculated critical exponents, $\beta \approx 1/17$ and $\delta \approx 35$, present a more complex picture. As shown in Table~\ref{tab:overview}, these values do not align with those of the regular 2D ($d=2$) or 3D ($d=3$) lattices. Intriguingly, the magnetization exponent $\beta \approx 1/17$ is even smaller than that of the 2D square lattice ($\beta = 1/8$), while the field response exponent $\delta \approx 35$ is significantly larger than the 2D value ($\delta=15$). This finding is significant as it disrupts any simple hierarchical ordering of critical exponents based solely on the Hausdorff dimension. For instance, a naive ordering by $d_H$ would place this fractal between the 2D and 3D regular lattices, but its exponents do not follow this interpolation.

\begin{table}[]
\begin{tabular}{|l|l|l|l|l|}
\hline
Geometry           & $d^{(\text{H})}$  & $d$                         & $\beta$     & $\delta$    \\ \hline
1D chain           & $1$                         & $1$                         & $1/\infty$ & $\infty$    \\ \hline
4x4 fractal~\cite{2dising, ad, acta}        & $\approx 1.792$ & $1.5$                       & $\approx 1/72$ & $\approx 205$  \\ \hline
SC(3, 1)~\cite{sierp}           & $\approx 1.893$             & $\approx 1.631$           & $\approx 1 / 7.4$ & *       \\ \hline
6x6 fractal~\cite{acta}        & $\approx 1.934$ & $\approx 1.774$ & $\approx 1/15$ & *                        \\ \hline
2D square lattice  & $2$                         & $2$                         & $1/8$                      & $15$                       \\ \hline
3D fractal (current study) & $2.5$     & $2$        & $\approx 1/17$ & $\approx 35$   \\ \hline
3D lattice~\cite{Xie2012, 3dising}         & $3$                         & $3$                         & $\approx 1/3$                      & $\approx 4.79$ \\ \hline
4D lattice         & $4$                         & $4$                         & $1/2$                      & $3$                        \\ \hline
\end{tabular}
\caption{
Critical exponents $\beta$ and $\delta$ of the classical Ising model with respect to dimensions. 
$d^{(\text{H})}$ is the Hausdorff dimension, whereas $d$ is the dimension derived from boundary scaling defined earlier. 
In the case of the 4x4 fractal, we report the values obtained by single-site measurements, rather than (partially) averaged or global results. 
In the case of the Sierpinski carpet SC(3, 1), we report the value of $\beta$ obtained by global measurement (as single-site measurements depend heavily on the location). 
3D lattice: $\beta \approx 0.3295$ by HOTRG~\cite{Xie2012}, $\beta \approx 0.3262$ by MC~\cite{3dising}.
Unknown values are replaced by star (*) symbol. 
} \label{tab:overview}
\end{table}

This leads to the central question in the study of critical phenomena on fractals: which dimension, if any, governs the universality class? Our results reinforce the notion that $d_H$ is not a sole predictor. The classical Ising model on the 2D Sierpiński gasket ($d_H \approx 1.585$) famously exhibits no finite-temperature phase transition \cite{Gefen1980Critical, Luscombe1985Critical}. This is often attributed to its finite order of ramification and its boundary scaling dimension $d=1$, which better reflects its quasi-one-dimensional nature at large scales. For the 3D fractal studied here, the boundary dimension is $d=2$, which, as discussed, appears consistent with the specific heat behavior but fails to determine the full universality class of the exponents.

For comparison, it is insightful to consider the transverse-field quantum Ising model, where the spatial dimension is effectively increased by one. As summarized in Table~\ref{tab:overview_quantum}, the critical exponents again do not follow a simple monotonic dependence on the fractal dimensions. A notable counterexample is the Sierpinski pyramid, which, despite having a Hausdorff dimension of $d_H=2$, exhibits a critical exponent $\beta = 0.25 \pm 0.02$ \cite{Yi2015QuantumIsingFractal}, a value distinct from the $\beta \approx 1/3$ of the regular 2D quantum Ising model (equivalent to the 3D classical Ising model).

\begin{table}[h]
  \begin{tabular}{|l|c|c|c|c|}
    \hline
    Geometry & $d^{(\text H)}$ & $d$ & $\beta$ & $\delta$ \\ \hline
    1D chain & 1 & 1 & $1/8$ & $15$ \\ \hline
    2D Sierpinski gasket~\cite{qsierp} & $\approx1.585$ & 1
        & $\approx 1/5$ & $\approx 8.7$ \\ \hline
    Sierpinski pyramid~\cite{Yi2015QuantumIsingFractal}
        & 2 & 1 & $0.25 \pm 0.02$ & ? \\ \hline
    2D lattice~\cite{Xie2012} & 2 & 2 & $\approx 1/3$ & $\approx 4.79$ \\ \hline
    3D lattice & 3 & 3 & $1 / 2$ & 3 \\ \hline
  \end{tabular}
  \caption{%
    Critical exponents $\beta$ and $\delta$ of the quantum Ising model in a transverse field.
    The universality classes correspond to the classical Ising model in one higher dimension.
  }
  \label{tab:overview_quantum}
\end{table}

Our results lend strong support to the idea that the concept of universality, while robust on regular lattices, applies only in a "weak" sense to fractal structures. This aligns with the argument, articulated by Carmona et al., that on self-similar structures the critical exponents are not determined by a single dimensional parameter but can depend on detailed geometrical properties like connectivity and lacunarity \cite{Carmona1998SierpinskiIsing}. The self-similar nature of the lattice means that such structural details at the smallest scale are magnified to the macroscopic scale, profoundly influencing the collective behavior at the critical point.

The results presented here are based on local measurements obtained by probing the system with an impurity tensor. A more complete picture could be obtained by calculating global quantities, for example, through automatic differentiation, though this approach is computationally prohibitive for a 3D system of this complexity with current methods. Future work should extend this analysis by investigating other 3D fractal lattices with different dimensional characteristics to disentangle the roles of $d_H$, $d$, and connectivity. Furthermore, determining the full set of critical exponents (e.g., $\alpha, \gamma, \nu$) would allow for a rigorous test of the validity of hyperscaling relations on this structure. Finally, exploring quantum phase transitions on this lattice, similar to studies on 2D Sierpiński fractals \cite{qsierp}, would be a fascinating avenue to understand how quantum fluctuations interact with this unique geometry.

\begin{acknowledgments}
The numerical calculations were performed on the K computer at the RIKEN Advanced Institute for Computational Science (AICS).
This work was partially supported 
1) by the EU NextGenerationEU through the Recovery and Resilience Plan for Slovakia under the project No.~09I03-03-V04-00682, 2) by the project IM-2021-26 (SUPERSPIN) funded by the Slovak Academy of Sciences via the programme IMPULZ 2021, and 3) by Agent\'{u}ra pre Podporu V\'{y}skumu a V\'{y}voja (No. APVV-20-0150) and Vedeck\'{a} Grantov\'{a} Agent\'{u}ra M\v{S}VVaM SR and SAV (VEGA No. 2/0156/22).
H.U. acknowledges support from JSPS KAKENHI Grant Numbers
JP21H05182 and JP21H05191, from MEXT Q-LEAP Grant
No. JPMXS0120319794, and from JST COI-NEXT No.
JPMJPF2014, and CREST No.JPMJCR24I1.
\end{acknowledgments}

\bibliography{refs}

\end{document}